\documentclass[12pt,a4,onecolumn,secnumarabic,amssymb, amsmath, nofootinbib,tightenlines,
nobibnotes, aps, prl,epsfig]{revtex4}
\usepackage{graphicx}
\usepackage{dcolumn}
\usepackage{bm}
\begin{document}
\preprint{APS/123-QED}
\title{ Analytical solution of the Longitudinal Structure Function $F_{L}$ in the Leading and Next- to- Leading- Order analysis at low $x$
with respect to Laguerre polynomials method}

\author{B.Rezaei }
\altaffiliation{rezaeibi@gmail.com}
\author{G.R.Boroun}
 \email{grboroun@gmail.com; boroun@razi.ac.ir}
\affiliation{ Physics Department, Razi University, Kermanshah
67149, Iran}
\date{\today}

\begin{abstract}
The aim of the present paper is to apply the Laguerre polynomials
method for the analytical solution of the Altarelli- Martinelli
equation. We use this method of the low $x$ gluon distribution to
the longitudinal structure function using MRST partons as input.
Having checked that this model gives a good description of the
data to predict of the longitudinal
structure function at leading and next to leading order analysis at low $x$.\\
\end{abstract}
 \pacs{11.55Jy, 12.38.-t, 14.70.Dj}
\keywords{Longitudinal structure function; Gluon distribution;
hard- Pomeron; Small-$x$; Regge- like behavior} 
\maketitle

The longitudinal structure function $F_{L}(x,Q^{2})$ comes as a
consequence of the violation of Callan- Gross relation [1] and is
defined as $F_{L}(x,Q^{2})=F_{2}(x,Q^{2})-2xF_{1}(x,Q^{2})$, where
$F_{2}(x,Q^{2})$ is the transverse structure function. As usual
$x$ is the Bjorken scaling parameter and $Q^{2}$ is the four
momentum transfer in a deep inelastic scattering process. In the
quark parton model (QPM) the structure function $F_{2}$ can be
expressed as a sum of the quark- antiquark momentum distributions
$xq_{i}(x)$ weighted with the square of the quark electric charges
$e_{i}$:
$F_{2}(x)={\sum_{i}}e_{i}^{2}x(q_{i}(x)+\overline{q}_{i}(x))$. For
spin $\frac{1}{2}$ partons QPM also predicts $F_{L}(x)=0$, which
leads to the Callan- Gross relation. This does not hold when the
quarks acquire transverse momenta from QCD radiation [2,3]. The
naive QPM has to be modified in QCD as quarks interact through
gluons, and can radiate gluons. Radiated gluons, in turn, can
split into quark- antiquark pairs (sea quarks) or gluons. The
gluon radiation results in a transverse momentum component of the
quarks. Thus, in QCD the longitudinal structure function is non-
zero. Due to its origin, $F_{L}$ is directly dependent to the
gluon distribution in the proton and therefore the measurement of
$F_{L}$ provides a sensitive test of perturbative QCD [4]. In this
way, the next- to- leading order (NLO) corrections to the
longitudinal structure function are large and negative, valid to
be at small
$x$ as shown at Refs.5-7.\\

As an illustration of this analysis, let us consider the Laguerre
polynomials method for solving the Altarelli- Martinelli
equation[8]. In recent years, Laguerre polynomials method have
proved to be valuable tools for the solving of the DGLAP [9]
equations by iterating the evolution over infinitesimal steps in
the fractional momentum $x$ [10,11]. This method yield numerical
solutions for DGLAP evolution equations. Here we use Laguerre
polynomials method, that it is useful for obtaining an analytical
solution to the longitudinal structure function in the Leading and
Next to Leading order analysis. The method is based on the search
for a solution in the form of a series and on decomposing of the
Altarelli- Martinelli equation kernels into a series in which the
terms are calculated recursively using the Laguerre polynomials
method.\\

I) Firstly, let us present a brief outline of the Laguerre
polynomials method in general. For this, the Laguerre polynomials
are defined as:
\begin{equation}
(n+1)L_{n+1}(x)=(2n+1-x)L_{n}(x)-nL_{n-1}(x),
\end{equation}
and orthogonality condition is defined as:
\begin{equation}
\int_{0}^{\infty}e^{-\acute{x}}L_{n}(\acute{x})L_{m}(\acute{x})d\acute{x}=\delta_{n,m}.
\end{equation}
The general integrable function $f(e^{-\acute{x}})$ is decomposed
into the sum:
\begin{equation}
f(e^{-\acute{x}})=\sum_{0}^{N}f(n)L_{n}(\acute{x}),
\end{equation}
where
\begin{equation}
f(n)=\int_{0}^{\infty}e^{-\acute{x}}L_{n}(\acute{x})f(e^{-\acute{x}})d\acute{x}.
\end{equation}\\

II) Secondly, we turn to the perturbative predictions for
$F_{L}(x,Q^{2})$, as QCD yields the Altarelli- Martinelli
equation. This equation can be written as [2,8,14]:
\begin{eqnarray}
x^{-1}F_{L}&=&<e^{2}>[\frac{\alpha_{s}}{4\pi}\textsf{c}_{L,g}^{LO}+(\frac{\alpha_{s}}{4\pi})^{2}\textsf{c}_{L,g}^{NLO}+...]{\otimes}
g+g{\rightarrow}q,
\end{eqnarray}
where the coefficient functions $\textsf{c}_{L,g}^{LO+NLO}$ for
$F_{L}(x,Q^{2})$ can be written as [12-16]:
\begin{equation}
\textsf{c}_{L,g}^{LO}(x)=8n_{f}x(1-x),
\end{equation}
and
\begin{equation}
\textsf{c}_{L,g}^{NLO}(x){\cong}n_{f}[(94.74-49.20x)x_{1}L_{1}^{2}+864.8x_{1}L_{1}+1161xL_{1}L_{0}+60.06xL_{0}^{2}+39.66x_{1}L_{0}-5.333(x^{-1}-1)],
\end{equation}
where used from these abbreviations [14],as:
\begin{equation}
x_{1}=1-x, L_{0}=Ln(x), L_{1}=Ln(x_{1})
\end{equation}
and $<e^{2}>=n_{f}^{-1}\sum_{i=1}^{N_{f}=4}e_{i}^{2}$ where
$<e^{2}>$ stand for the average of the charge $e^{2}$ for the
active quark flavours.\\

Now, Eq.(5) shows expliciting the dependence  of $F_{L}$ on the
strong coupling constant and the gluon density, as at small $x$
the gluon distribution is the dominant one. Thus the gluonic
contribution $F_{L}^{g}$ to $F_{L}$ in Eq.(5) is reduced to:
\begin{eqnarray}
x^{-1}F_{L}^{g}&=&<e^{2}>[\frac{\alpha_{s}}{4\pi}\textsf{c}_{L,g}^{LO}+(\frac{\alpha_{s}}{4\pi})^{2}\textsf{c}_{L,g}^{NLO}+...]{\otimes}
g.
\end{eqnarray}
The running coupling constant $\alpha_{s}(Q^{2})$ has the
approximate analytical form in NLO:
\begin{equation}
\frac{\alpha_{s}(Q^{2})}{2\pi}=\frac{2}{\beta_{0}\ln(\frac{Q^{2}}{\Lambda^{2}})}
[1-\frac{\beta_{1}\ln\ln(\frac{Q^{2}}{\Lambda^{2}})}{\beta_{0}^{2}\ln(\frac{Q^{2}}{\Lambda^{2}})}],
\end{equation}
where  $\beta_{0}=\frac{1}{3}(33-2N_{f})$ and
$\beta_{1}=102-\frac{38}{3}N_{f}$ are the one- loop (LO) and the
two- loop (NLO) correction to the QCD $\beta$- function, $N_{f}$
being the number of active quark flavours ($N_{f}=4$). In Ref.[17]
the authors have suggested that expression (9) at leading order
can be reasonably approximated by
$F_{L}(x,Q^{2}){\approx}0.3\frac{4\alpha_{s}}{3\pi}xg(2.5x,Q^{2})$,
which demonstrates the close relation between the longitudinal
structure
function and the gluon distribution.\\

III) Thirdly, In what follows we calculate $F_{L}^{g}$ using the
the Laguerre polynomials method. We used the variable
transformation, $x=e^{-x'}$ and $y=e^{-y'}$ to get from the
Altarelli- Martinelli equation form to the Laguerre polynomials
form. Next, we combine and expand each terms of this equation on
Laguerre polynomials according to relations (3) and (4) and used
this properties as;
$\int_{0}^{x'}L_{n}(x'-y')L_{m}(y')dy'=L_{n+m}(x')-L_{n+m+1}(x')$,
we find an equation which determines $F_{L}^{g}(x,Q^{2})$ in terms
of Laguerre polynomials:
\begin{equation}
\sum_{n=0}^{N}F^{g}_{Ln}(n,Q^{2})L_{n}(x')=\sum_{n_{1},n_{2}}{K}_{L,g}(n_{1},Q^{2})G_{n_{2}}(n_{2},Q^{2})[L_{n_{1}+n_{1}}(x')-L_{n_{1}+n_{1}+1}(x')],
\end{equation}
or
\begin{equation}
F^{g}_{Ln}(n,Q^{2})=\sum_{m=0}^{n}G(m,Q^{2})[{K}_{L,g}(n-m)-{K}_{L,g}(n-m-1)],
\end{equation}
 where
 ${K}_{L,g}(n)=\frac{\alpha_{s}}{4\pi}\textsf{c}_{L,g}^{LO}(n)+(\frac{\alpha_{s}}{4\pi})^{2}\textsf{c}_{L,g}^{NLO}(n)$ and
 $
G(n,Q^{2})=\int_{0}^{\infty}e^{-\acute{x}}L_{n}(\acute{x})G(e^{-\acute{x}},Q^{2})d\acute{x}
$. Therefore we find the solution of the longitudinal structure
function defined by solving this recursion relation, as:
\begin{equation}
F^{g}_{L}(x,Q^{2})=\sum_{n=0}^{N}F^{g}_{Ln}(n,Q^{2})L_{n}(Ln\frac{1}{x}).\\
\end{equation}
This result is completely general and gives the LO and NLO
expression for the longitudinal structure function once the gluon
distribution is known with help of other standard gluon
distribution function [15-16,18-23]. Here we can expand the
integrable functions till a finite order $N=30$, as we can
convergence these
series in the numerical determinations.\\

 We computed the predictions for all detail of the longitudinal
structure function in the kinematic range where it has been
measured by $H1$ collaboration [24-27] and compared with DL model
[22] based on hard Pomeron exchange, also compared with
computation Moch, Vermaseren and Vogt [14-16] at the second order
with input data from MRST and also with Block model [23]. Our
numerical predictions are presented as functions of $x$ for the
$Q^{2}=$20 $GeV^{2}$. The average value $\Lambda$ in our
calculations  is corresponding to ${\simeq}0.22
 GeV$ at LO and corresponding to ${\simeq}0.323
 GeV$ at NLO [19]. The
results are presented in Fig.1 where they are compared with the
very recent $H1$ data [27] and with the results obtained with the
help of other standard gluon distribution functions.\\

The curves represent the LO and NLO QCD calculations
 $F_{L}$ based on a fit to all data. We compare our results with
 predictions of $F_{L}$ up to NLO in perturbative QCD  that
 the input densities is given by MRST parameterizations [19]. Also, we compare our results
  with and without the next-to-leading-order corrections with the two pomeron fit as is
seen in Fig.1. We see immediately that the next-to-leading-order
corrections have opposite signs for the standard gluons. To
emphasize the size of the next-to-leading-order corrections, we
show in Fig.2 the ratio (LO+NLO/LO) and compared with respect to
MRST gluon distribution, at $Q^{2}=$20 $GeV^{2}$. The agreement
between the Laguerre polynomials method and data is remarkably
good. The good agreement indicates that the Laguerre polynomials
method has a good asymptotic behavior and it is compatible both
with the data and with the other standard models. These results
indicate that the Laguerre polynomials method is a good method to
solve the Altarelli- Martinelli equation for the longitudinal
structure function at LO and NLO analysis. As this model has this
advantage that we get a very elegant solution for the longitudinal
structure function. In this case, we will be able to verify the
results between the longitudinal structure function and the gluon
distribution function at the same $x$ point, and these results
extend our knowledge about of the longitudinal structure function
into the
   low- $x$ region.\\

In summary, we have used the Laguerre polynomials method  for low
$x$ the gluonic contribution to the longitudinal structure
function, slightly changing the parameters fixed from previous
analysis, to fit HERA data on $F_{L}$. And also we have obtained
an analytic solution for the longitudinal structure function in
the next- to- leading order at low $x$. Having checked that this
model gives a good description of the data, we have used it to
predict $F_{L}$ to be measured in electron- proton collisions. The
results are close to those obtained with other models. The
conclusion of this exercise is that the Laguerre polynomials
method, simple as it is, and has the short time consuming on the
numerical calculations as it is a real advantage to realize fits
to PQCD. To confirm the method
 and results, the calculated values are compared with the $H1$ data on the longitudinal
 structure function, at small $x$ and QCD fits.\\

\textbf{References}\\
\hspace{2cm}1.G.G.Callan and D.Gross, Phys.Lett.B\textbf{22}, 156(1969).\\
\hspace{2cm}2.F.Carvalho, et.al. Phys.Rev.C\textbf{79}, 035211(2009).\\
\hspace{2cm}3.V.P.Goncalves and
M.V.T.machado,Eur.Phys.J.C\textbf{37}, 299(2004);
M.V.T.machado,Eur.Phys.J.C\textbf{47}, 365(2006).\\
\hspace{2cm}4. R.G.Roberts, The structure of the proton, (Cambridge University Press 1990)Cambridge.\\
\hspace{2cm}5. A.V.Kotikov, JETP Lett.\textbf{59}, 1(1994); Phys.Lett.B\textbf{338}, 349(1994).\\
\hspace{2cm}6. Yu.L.Dokshitzer, D.V.Shirkov, Z.Phys.C\textbf{67},
449(1995);W.K.Wong, Phys.Rev.D\textbf{54}, 1094(1996).\\
\hspace{2cm}7. G.R.Boroun, International Journal of Modern Physics
E, Vol.18, No.1, 131(2009).\\
\hspace{2cm}8. G.Altarelli and G.Martinelli, Phys.Lett.B\textbf{76}, 89(1978).\\
\hspace{2cm}9. Yu.L.Dokshitzer, Sov.Phys.JETP {\textbf{46}},
641(1977); G.Altarelli and G.Parisi, Nucl.Phys.B \textbf{126},
298(1977); V.N.Gribov and L.N.Lipatov,
Sov.J.Nucl.Phys. \textbf{15}, 438(1972).\\
10. L.Schoeffel, Nucl. Instrum.Math.A{\textbf{423}},
439(1999).\\
11. W.Furmanski and R.Petronzio, Nucl.Phys.B{\textbf{195}}, 237(1982).\\
12. J.L.Miramontes, J.sanchez Guillen and E.Zas, Phys.Rev.D \textbf{35}, 863(1987).\\
13. D.I.Kazakov, et.al., Phys.Rev.Lett. \textbf{65}, 1535(1990).\\
14. S.Moch, J.A.M.Vermaseren, A.vogt, Phys.Lett.B \textbf{606},
123(2005).\\
15. A.D.Martin, W.J.Stirling,R.Thorne, Phys.Lett.B \textbf{635}, 305(2006).\\
16. A.D.Martin, W.J.Stirling,R.Thorne, Phys.Lett.B \textbf{636}, 259(2006).\\
17. A.M.Cooper-Sarkar et.al., Z.Phys.C\textbf{39},
 281(1998); A.M.Cooper-Sarkar and R.C.E.Devenish, Acta.Phys.Polon.B\textbf{34},
 2911(2003).\\
18. A.D.Martin, W.S.Striling and R.G.Roberts, Euro.J.Phys.C {\bf 23}, 73(2002).\\
19. A.D.Martin, R.G.Roberts, W.J.Stirling,R.Thorne, Phys.Lett.B \textbf{531}, 216(2002).\\
20. M.Gluk, E.Reya and A.Vogt, Euro.J.Phys.C\textbf{5}, 461(1998).\\
21. A.Vogt, S.Moch, J.A.M.Vermaseren, Nucl.Phys.B \textbf{691},
129(2004).\\
22. A. Donnachie and P.V.Landshoff, Phys.Lett.B\textbf{533},
277(2002); Phys.Lett.B\textbf{550}, 160(2002);\\ J.R.Cudell, A.
Donnachie and P.V.Landshoff, Phys.Lett.B\textbf{448}, 281(1999);\\
P.V.Landshoff, hep-ph/0203084.\\
23. M.M.Block et.al., Phys.Rev.D\textbf{77}, 094003(2008).\\
24. S.Aid et.al, $H1$ collab. phys.Lett.B {\textbf 393}, 452(1997).\\
25. C.Adloff et.al, $H{1}$ Collab., Eur.Phys.J.C\textbf{21}, 33(2001); phys.Lett.B {\bf 393}, 452(1997).\\
26. N.Gogitidze et.al, $H{1}$ Collab., J.Phys.G\textbf{28}, 751(2002).\\
27. F.D.Aaron, $H1$ collab. phys.Lett.B {\textbf 665}, 139 (2008).\\

\subsection{Figure captions }
Fig 1:The values of the gluonic contribution to the longitudinal
structure function at $Q^{2}=20\hspace{0.1cm} GeV^{2}$ in LO and
NLO analysis by solving the Altarelli- Martinelli equation with
respect to Laguerre polynomials method
  that compared with H1 Collab. data
    (up and down triangle). The error on the  H1
 data is the total uncertainty of the determination of
 $F_{L}$ representing the statistical, the systematic and the model errors added in quadrature.
Circle data are the MVV prediction [20 ]. The solid line is the
NLO QCD fit to the H1 data for $y<0.35$ and
  $Q^{2}{\geq}3.5\hspace{0.1cm}GeV^{2}$.  The dot line is the DL fit [22] and the dash line is a
   QCD fit with respect to LO gluon distribution function from Block [23] analysis.\\

   Fig 2:The ratio of the next-to-leading to leading-order, the next-to-leading to MRST gluonic distribution and the leading-order to MRST predictions
   for the gluonic distribution to the longitudinal structure function with respect to the Laguerre polynomials
   method at $Q^{2}=20\hspace{0.1cm} GeV^{2}$.\\
\begin{figure}
\includegraphics[width=1\textwidth]{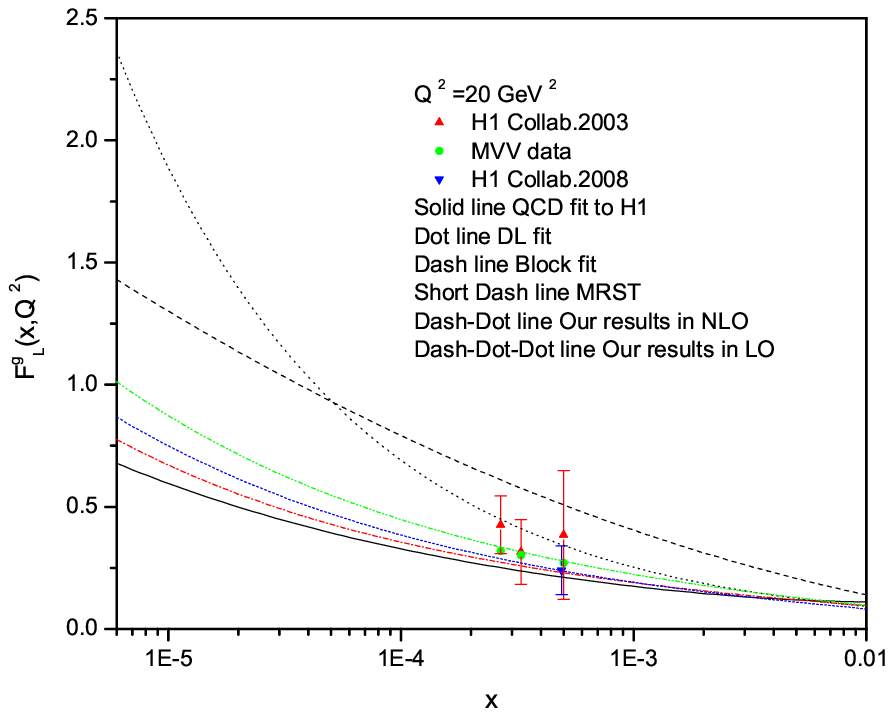}
\caption{} \label{Fig11}
\end{figure}
\begin{figure}
\includegraphics[width=1\textwidth]{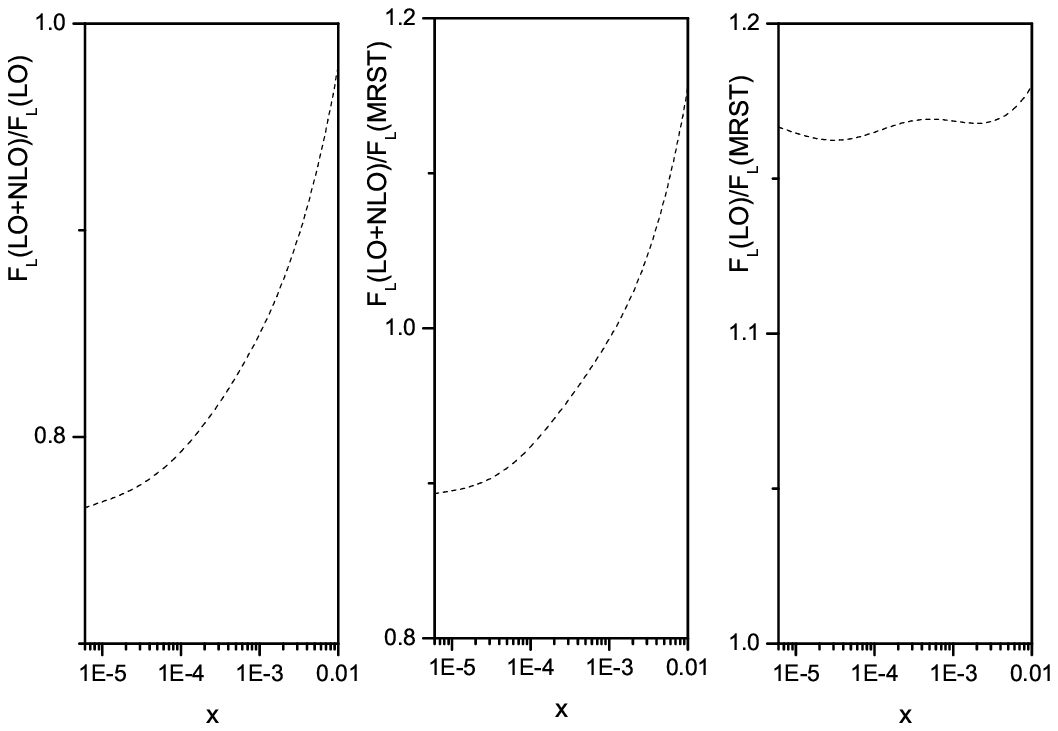}
\caption{} \label{Fig2}
\end{figure}

\end{document}